\documentclass[%
 reprint,
showpacs,preprintnumbers,
amsmath,amssymb,
aps,
prb,
longbibliography
]{revtex4-1}

\usepackage{epsfig}
\usepackage{array}
\usepackage{multirow}
\usepackage{amsmath}
\usepackage{braket}
\usepackage{graphicx}
\begin{document}

\title{Momentum-dependent spin selection rule in photoemission with glide symmetry}

\author{Ji Hoon Ryoo}
\author{Cheol-Hwan Park}
\email{cheolhwan@snu.ac.kr}
\affiliation{Department of Physics, Seoul National University, Seoul 08826, Korea}

\date{\today}

\begin{abstract}

We present a comprehensive theory on the spin- and angle-resolved photoemission spectroscopy (SARPES) of materials with glide-mirror symmetry, focusing on the role of glide symmetry on the spin selection rule.
In the glide-symmetric SARPES configuration, where the surface of a material, the incoming light and the outgoing photoelectrons are invariant under a glide reflection, the spin polarization of photoelectrons is determined by the glide eigenvalue of the initial state, which makes SARPES a powerful tool for studying topological phases protected by glide symmetry.
We also show that, due to the nonsymmorphic character of glide symmetry,
the spin polarization of a photoelectron whose momentum is in the second surface Brillouin zone is the opposite of the spin polarization of a photoelectron which is ejected from the same initial Bloch state but whose momentum is in the first zone.
This momentum dependence of spin selection rule clearly distinguishes glide symmetry from mirror symmetry and is particularly important if the Bloch wavevector of the initial state is close to the first surface Brillouin zone boundary. As a proof of principle, we simulate the SARPES from the surface states of KHgSb (010) and investigate how the spin selection rule imposed by the glide symmetry manifests itself in a real material.

\end{abstract}

\maketitle

\section{introduction}

In study of a crystal possessing a certain set of symmetries,
the electronic band energies and wavefunctions 
at symmetry-invariant crystal momenta
provide valuable information on the electronic structure over the whole Brillouin zone.
The representation of wavefunctions at a symmetry-invariant $k$ point alone
greatly restricts the form of
the electronic structure at all the nearby $k$-points.
For example, the effective Hamiltonian of the $\textrm{Bi}_2\textrm{Se}_3$ (111) surface is nearly isotropic with respect to the crystal momentum $\mathbf{k}$ around $\Gamma$~\cite{park2012spin, zhang2013spin}
with its leading correction proportional to $k^3$ yielding the hexagonal warping of the Fermi circle~\cite{fu2009hexagonal} thanks to the $C_{3v}$ symmetry at $\Gamma$.
In addition to the local information near the high-symmetry points or lines, analysis of symmetry representations at symmetry-invariant $k$ points
reveals topological aspects of the electronic structure throughout the entire Brillouin zone.
For example, a glide-symmetric two-dimensional material hosts at least one Dirac point along the glide-symmetric line of the Brillouin zone regardless of the strength of the spin-orbit coupling (SOC)~\cite{young2015dirac}. Also,
the three-dimensional Dirac point of $\textrm{Na}_3\textrm{Bi}$ which resides on its fourfold rotation axis is robust against any deformation of the crystal structure as long as the symmetry is preserved~\cite{Liu2014discovery}.

On the other hand,
simulating the photoelectron intensity in angle-resolved photoemission spectroscopy (ARPES) from first principles is quite subtle
due to the effects of elastic multiple scattering by ions, inelastic scattering by various collective excitations in a solid (and the finite inelastic lifetime and the inner potential thereby induced), and the change in the electric field of light when the light penetrates the surface of a material.
In particular, neglecting the multiple scattering effect by the lattice potential and
assuming a photoelectron state to be a plane-wave state even inside the material
may often fail to provide accurate values of photocurrent intensities which are necessary to simulate the circular or linear dichroism and the spin polarization of photoelectrons from first principles.

As an important example, if we assume that final states in a photoemission process are plane-wave states, then the dipole matrix element
$\langle \mathbf{k}^f | \mathbf{A}\cdot \mathbf{p} | i \rangle = \mathbf{A}\cdot\mathbf{k}^f \langle \mathbf{k}^f | i \rangle$
vanishes when light polarization $\mathbf{A}$ and the momentum of the photoelectron $\mathbf{k}^f$ are perpendicular to each other.
Here, $|i\rangle$ is the initial electronic state and $|\mathbf{k}^f\rangle$ the (plane-wave) final state.
Similarly, if we assume that final states in spin- and angle-resolved photoemission spectroscopy (SARPES) are the direct product of the plane wave $|\mathbf{k}^f\rangle$ and a constant spinor $|\sigma^i\rangle$ ($\sigma^i=\pm 1$), the ratio between the spin-up and spin-down photoelectron intensities would be $\langle \mathbf{k}^f, 1 | i \rangle / \langle\mathbf{k}^f,-1 | i \rangle$,
which cannot explain the dependence of the spin polarization of photoelectrons on $\mathbf{A}$ observed in experiments~\cite{jozwiak2013photoelectron}.
To avoid this difficulty while retaining the simplicity of the plane-wave approximation for the final states,
one may assume that the relevant part of the final state in photoemission processes is the Bloch sum of the atomic orbitals whose phase factor is similar to the plane wave~\cite{Damascelli2013PRL,Damascelli2014PRL}.
This approach yields the light-polarization dependence of the spin polarization of photoelectrons which was absent in the simple plane-wave approximation~\cite{Ryoo2016PRB,Hwang2011PRB}.
Still, however, the accurate description of the photon-energy dependence of the ARPES intensities requires a better treatment of final states.

Remarkably, the analysis of ARPES at symmetry-invariant $k$ points can bypass this difficulty and provide useful guidance for interpreting ARPES.
The symmetry group of the ARPES configuration including the direction of the light polarization, the momentum of photoelectrons, and the surface of a solid is smaller than the symmetry group of the surface alone in general.
However, when the symmetry group of the ARPES configuration is the same as that of the surface,
ARPES reflects the rich information on the underlying electronic structure~\cite{Kobayashi2017PRB,yaji2017spin}
and a symmetry analysis provides exact results on ARPES without subtle assumptions on the final state of photoelectrons.
For example, owing to the mirror symmetry,
when the polarization of the light lies parallel to (perpendicular to) the mirror plane, the spin polarization of the photoelectrons must be parallel (anti-parallel) to that of the surface state of $\textrm{Bi}_2\textrm{Se}_3$ (111) \cite{Ryoo2016PRB, Gotlieb2017PRB}.
As another example, the absence of the circular dichroism on the mirror-invariant line in the Brillouin zone of a superconductor $\textrm{Bi}_2\textrm{Sr}_2\textrm{Ca}\textrm{Cu}_2\textrm{O}_{8+\delta}$ plays an important role in proving the nonexistence of an order which breaks the mirror symmetry~\cite{he2016angle}.

Symmetry also gives valuable information on ARPES configuration where the symmetry is broken. Consider the ARPES on $\textrm{Bi}_2\textrm{Se}_3$ (111) where the directions of propagation of light and photoelectrons lie in the mirror plane but the light polarization is arbitrary. In this case, the spin polarization of the photoelectrons as a function of the light polarization is completely determined by symmetry, apart from one complex-valued parameter which can be obtained from experiment~\cite{Kuroda2016PRB,Kobayashi2017PRB}. Therefore, symmetry analysis gives both qualitative and quantitative results in ARPES configuration with or without symmetry.

So far we have discussed a few cases where mirror symmetry plays a key role in interpreting ARPES experiments. In addition, the implication of the glide-mirror symmetry to ARPES, when SOC is negligible, has been studied in several papers~\cite{pescia1985determination, prince1986symmetry, Arpiainen2006PRL_BSCCO}.
Due to the fractional translation contained in a glide symmetry operation, the ARPES selection rule imposed by glide symmetry depends on the momentum of photoelectrons.
Suppose that the light polarization is perpendicular or parallel to the glide plane and 
translation and glide symmetry allows the ejection of an electron from a certain valence state to a photoelectron state whose momentum is in the first surface Brillouin zone. Then, the transition from the same initial state induced by the same light is forbidden by symmetry when the momentum of the photoelectron is in the second surface Brillouin zone~\cite{pescia1985determination, prince1986symmetry, Arpiainen2006PRL_BSCCO}.

On the other hand, in glide-invariant systems of recent interest, spin degrees of freedom play an important role in the electronic structure.
For example, a broad range of glide-invariant topological semimetals are predicted to possess a line node within theories neglecting SOC,
but SOC induces a gap on the band-crossing line except at a finite number of $k$-points,
thus reducing the line node to point nodes
~\cite{Fu2015PRL,bzduvsek2016nodal,shao2018nonsymmorphic}.
Also, even in glide-invariant materials with weak SOC, such as black phosphorus,
the effect of SOC becomes important when heavy atoms are adsorbed to the surface~\cite{Ehlen2018PRB}.
Despite the abundance of materials with strong SOC which are symmetric under a glide operation, however, the optical selection rule for those materials imposed by glide symmetry has not yet been investigated so far.

The optical selection rule in spinless glide-invariant systems is not directly applicable in systems with strong SOC.
When we take into account the effect of the spin degree of freedom, the number of the photoelectron states at a given energy $E^f$ and the momentum $\mathbf{k}^f$ are doubled (i.e., there are final states with two opposite spin directions in the asymptotic vacuum with \textit{different} glide eigenvalues).
Therefore, contrary to the spinless case, the electronic bands observed in the first surface Brillouin zone of the photoelectrons in spin-integrated ARPES should also be observed in the second zone.

In this paper, we report the implication of glide symmetry to the photoemission from glide-symmetric surfaces or two-dimensional materials.
Especially we show that, due to the nonsymmorphic character of the glide reflection, the spin polarization of the photoelectron whose momentum is in the first surface Brillouin zone is the opposite of the spin polarization of the photoelectron in the second zone ejected from the same initial state.
This momentum dependence in the selection rule imposed by glide symmetry is absent in the mirror-symmetric case, and offers an experimental method to differentiate a glide plane from a mirror plane.
Our paper develops a unified and comprehensive theory to explain the photoelectric effect from materials with mirror or glide symmetry.
For demonstration purposes, we apply our theory to the SARPES from the so-called hourglass surface states of KHgSb (010)~\cite{wang2016hourglass,ezawa2016PRB}.

%%%%%%%%%%%%%%%%%%%%%%%%%%%
%%%%%%%%%%%%%%%%%%%%%%%%%%%

\section{Methods}

We investigated the photoemission process from a non-degenerate Bloch state $|i\rangle$ with crystal momentum $\mathbf{k}^i$ to an outgoing photoelectron state $|f,\sigma\hat{y}\rangle$ with momentum $\mathbf{k}^f$, energy $E^f$ and the spin quantum number $\sigma/2$ $(\sigma=\pm 1)$ along the $y$ axis in the asymptotic vacuum.
We suppose that a material (whose surface lies in the $zx$ plane) is invariant under glide reflection $(x,y,z)\mapsto (x,-y,z+c/2)$,
where $c$ is the lattice parameter along the translational direction of the glide reflection.
Since we are concerned with the implication of the glide symmetry to photoemission, we assume that $\mathbf{k}^i$ and $\mathbf{k}^f$ are \textit{invariant under the glide reflection} (Fig.~2b); thus, $k_y^i=k_y^f=0$. (Especially, we will only refer to the line $k_y^i=0$ in the surface Brillouin zone as the glide-invariant line hereafter and will not consider states with $k_y^i=\pi/c$, which yield photoelectrons breaking the glide symmetry.)
Due to the lattice symmetry, the in-plane components of the initial and final states, $\mathbf{k}_{\|}^i=(0,0,k_z^i)$ and $\mathbf{k}_{\|}^f=(0,0,k_z^f)$, satisfy
\begin{equation}
\mathbf{k}_{\|}^f-\mathbf{k}_{\|}^i=\mathbf{G}_{\|}=n(2\pi/c)\hat{z},
\label{eq:kkg}
\end{equation}
where $\mathbf{G}_{\|}$ is a surface reciprocal lattice vector parallel to the glide plane and $n$ is an integer.

We used the dipole approximation to describe the photoemission.
Then the spin polarization of the photoelectron is given by 
$\mathbf{P}^f=\langle\chi|2\mathbf{S}|\chi\rangle/\langle\chi|\chi\rangle$
with
%\begin{equation}
$
|\chi\rangle=\sum_\sigma{|\sigma\hat{y}\rangle \langle f,\sigma\hat{y}|\mathbf{A}\cdot\mathbf{p}|i\rangle },
$
%\end{equation}
  where $\mathbf{S}$ denotes the spin operator in unit of $\hbar$, $\mathbf{A}$ a constant vector parallel to the light polarization, and $|\sigma\hat{y}\rangle$ the constant two-component spinor fully polarized along $\sigma\hat{y}$~\cite{park2012spin}.

After deriving the SARPES selection rule imposed by the glide symmetry, we numerically demonstrated the validity of the selection rule by simulating the SARPES of the surface bands of KHgSb (010).
In order to simulate the electronic structure of KHgSb, we used an \textit{ab initio} tight binding method. We used Quantum Espresso package for density-functional-theory (DFT) calculations of the bulk material~\cite{giannozzi2009quantum} with the PBEsol functional~\cite{PerdewPRL2008} for the exchange-correlation energy. The energy cutoff for wavefunctions was set to 80 Ry and the Brillouin zone was sampled on a  uniform $8\times 8 \times 8$ grid.

We constructed maximally localized Wannier functions which accurately describe the bands near the band gap using Wannier90 package~\cite{mostofi2014updated}. From those Wannier functions, which mainly consist of Hg $6s$ orbitals and Sb $5p$ orbitals, we extracted the hopping integrals of the tight-binding model for the surface calculation.

%%%%%%%%%%%%%%%%%%%%%%%%%%%
%%%%%%%%%%%%%%%%%%%%%%%%%%%

\section {Results and discussion}

Let $\bar{M}_y$ be the glide operation which acts on a spinor as $e^{i\pi S_y}$. Since $\mathbf{k}^i$ is on the glide-invariant line, the initial state is an eigenstate of $\bar{M}_y$:
\begin{equation}
\bar{M}_y |i\rangle=i \lambda \exp{[-ik_z^i c/2]}|i\rangle\quad(\lambda=\pm 1).
\label{eq:eig}
\end{equation}
As is well-known, compared with the eigenvalue of an ordinary mirror operation, the eigenvalue of the glide reflection contains an additional phase factor $\exp{[-ik_z^i c/2]}$ which depends on $\mathbf{k}^i$.

Similarly, the final state $|f,\sigma\hat{y}\rangle$ is also an eigenstate of $\bar{M}_y$. Since the final state in the asymptotic vacuum is the product of a plane wave $\exp{[i\mathbf{k}^f\cdot\mathbf{r}]}$ and a constant spinor $|\sigma\hat{y}\rangle$ \textit{no matter how complicated the wavefunction is near or inside the crystal}, the eigenvalue of $|f,\sigma\hat{y}\rangle$ with respect to the glide symmetry is $i\sigma \exp{[-ik_z^f c/2]}$.

When the light is $p$-polarized, i.e., the light polarization is parallel to the glide plane, $\bar{M}_y(\mathbf{A}\cdot\mathbf{p})\bar{M}_y^{-1}=\mathbf{A}\cdot\mathbf{p}$. Therefore, in this case, the photoemission is allowed only if the glide eigenvalues of the initial and final states are equal:
\begin{equation}
i\lambda\exp{[-ik_z^i c/2]} = i\sigma\exp{[-ik_z^f c/2]}.
\label{lambda_sigma_eig}
\end{equation}
Using Eq.~(\ref{eq:kkg}), we see that the photoemission is allowed only if
\begin{equation}
\sigma=(-1)^n \lambda\quad(p\textrm{-polarized light}).
\label{lambda_sigma_p}
\end{equation}
Hence, the photoelectrons ejected by $p$-polarized light from the initial state with glide eigenvalue $i\lambda\exp{[-ik_z^i c/2]}$ is fully spin-polarized along $\sigma\hat{y}=(-1)^n\lambda\hat{y}$.
Similarly, when the light is $s$-polarized i.e., the light polarization is perpendicular to the glide plane, $\bar{M}_y(\mathbf{A}\cdot\mathbf{p})\bar{M}_y^{-1}=-\mathbf{A}\cdot\mathbf{p}$ and thus the photoemission is allowed only if
\begin{equation}
\sigma=(-1)^{n+1} \lambda\quad(s\textrm{-polarized light}).
\label{lambda_sigma_s}
\end{equation}
Thus the photoelectrons ejected by $p$- and $s$-polarized light are fully spin-polarized in the opposite directions.
Equations~(\ref{lambda_sigma_eig}), (\ref{lambda_sigma_p}) and (\ref{lambda_sigma_s}) are the key results of our paper.

The selection rule derived above for a glide-symmetric configuration is clearly different from the selection rule for a mirror-symmetric configuration.
If a system is invariant with respect to ordinary mirror reflection $(x,y,z)\mapsto(x,-y,z)$ instead of glide reflection,
then the mirror eigenvalues of the initial and final states are $i\lambda$ and $i\sigma$, respectively, which are $\mathbf{k}$-independent.
Therefore, $p$-polarized light allows the transition with $\sigma=\lambda$ and $s$-polarized light allows $\sigma=-\lambda$ \textit{regardless of} $\mathbf{k}^i$ \textit{or} $\mathbf{k}^f$.
%% Table
However, in the glide-symmetric configuration, the glide eigenvalue of the initial state has a $\mathbf{k}^i$-dependent phase factor
while the eigenvalue of the final state contains a $\mathbf{k}^f$-dependent factor.
These two different phase factors do not cancel out, introducing an additional element
$\exp{[i(k_z^f-k_z^i)c/2]}$ to the selection rule.
Therefore, in the glide-symmetric configuration, the photoelectron with momentum $(\sqrt{2mE^f/\hbar^2-(k_z^f)^2},0,k_z^f)$ and the photoelectron with momentum $(\sqrt{2mE^f/\hbar^2-(k_z^f+2\pi/c)^2},0,k_z^f+2\pi/c)$ ejected from the same initial surface state $|i\rangle$ by the same light are fully spin-polarized in the directions opposite to each other.

  \begin{figure}
  \centering
  \includegraphics[width=0.92\columnwidth]{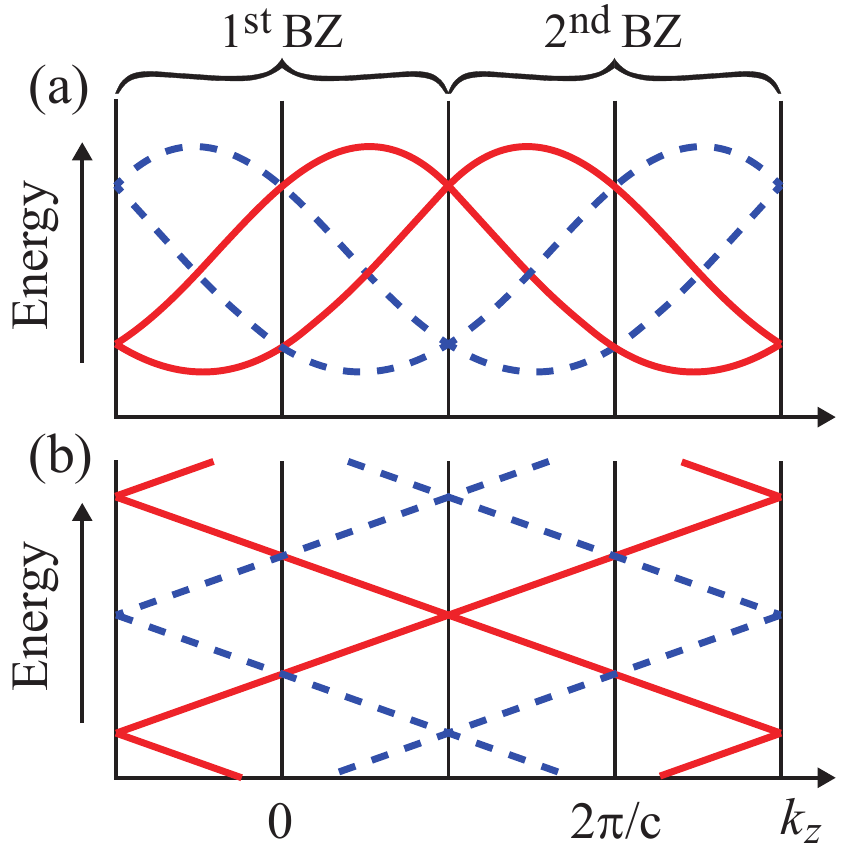}
  \caption{
  A schematic of the SARPES data with the glide-symmetric configuration, where the light is $p$-polarized. Solid red curves show the energy bands of a two-dimensional material or the surface energy bands of a bulk material which yield photoelectrons whose spin is fully polarized in a direction perpendicular to the glide plane (say, spin up), and dashed blue curves show the bands which yield photoelectrons whose spin is fully polarized in the opposite direction (spin down). (a) and (b) show two types of possible band connectivity.
  }
  \label{sarpes}
  \end{figure}
  
We remark that our theory on the glide selection rule is valid for any non-degenerate initial states of a glide-symmetric surface, regardless of the time-reversal symmetry. In the case of a two-dimensional glide-symmetric material,
the theory is applicable except only when the material is invariant under $PT$, the combination of the spatial inversion and the time reversal, which would make every energy band doubly degenerate. Despite its generality, we demonstrate our theory using non-magnetic materials.

Figure~1 shows a typical behavior of the spin polarization of photoelectrons from SARPES  on the glide-invariant line of the surface Brillouin zone when the light is $p$-polarized, assuming the material is non-magnetic.
Since the time-reversal and glide-reflection operators commute, every Kramer pair at $k_z^i=0$ or $\pi/c$ is composed of two states the glide eigenvalues of which are complex conjugates of each other.
Therefore, the glide eigenvalues of a Kramer pair [Eq.~(\ref{eq:eig})] at $k_z^i=0$
must be different from each other ($\pm i$) while at $k_z^i=\pi/c$ they must be the same, either $1$ or $-1$.
This fact implies that, in Fig.~1, two bands connected to a single Kramer pair at $k_z=0$ have different glide eigenvalues and thus the photoelectrons ejected from those two bands are spin-polarized in opposite directions. Near $k_z^i=\pi/c$, on the other hand, two bands which are degenerate at $k_z=\pi/c$ have the same glide eigenvalue and the spin polarization of the photoelectrons ejected from those two bands are the same.

More importantly, due to the dependence of the glide eigenvalue on the momentum of photoelectrons [Eq.~(\ref{eq:eig})], the spin polarization of a photoelectron with its in-plane momentum in the second surface Brillouin zone is the opposite of the counterpart in the first zone even if those electrons are ejected from the same initial state (Fig.~1). In particular, the photoelectrons at two first-zone boundaries ($k_z^f=\pm \pi/c$) are spin-polarized in opposite directions.
On the contrary, in the mirror-symmetric case, the photoelectrons with different in-plane momenta ejected from a single initial state always have the same spin polarization. Table~\ref{tab:summary} summarizes the result.

\begin{table}
\renewcommand{\arraystretch}{1.5}
\begin{tabular}{m{2.5 cm}| >{\centering\arraybackslash} m{2.3cm} | >{\centering\arraybackslash} m{2.3cm}}
\hline
& $p$-pol. light & $s$-pol. light \\
\hline
\multicolumn{3}{l}{Glide-symmetric: $\bar{M}_y|i\rangle=i\lambda e^{-ik_z^i c/2}|i\rangle$}  \\
\hline
Even $n$ & $\sigma=\lambda$ & $\sigma=-\lambda$ \\
Odd $n$ & $\sigma=-\lambda$ & $\sigma=\lambda$ \\
\hline
\hline
\multicolumn{3}{l}{Mirror-symmetric: $M_y|i\rangle=i\lambda|i\rangle$}  \\
\hline
Any $n$ & $\sigma=\lambda$ & $\sigma=-\lambda$ \\
\hline
\end{tabular}
\caption{The selection rules determining the spin polarization of photoelectrons. Here $n$ is the integer satisfying $\mathbf{k}^f_{\|}-\mathbf{k}_{\|}^i=n(2\pi/c)\hat{z}$ [see Eq.~(\ref{eq:kkg})] and $\sigma$ denotes the spin of the final state along $\hat{y}$.}
\label{tab:summary}
\end{table}

The difference in the SARPES behaviors between glide-symmetric and mirror-symmetric configurations is quite remarkable
since
many physical quantities such as electrical conductivity, Raman tensor, and stiffness tensor in elasticity theory
depend only on the point group of the crystal rather than the full space group~\cite{Kleiner1966PR,comment_kubo}.
In SARPES, however, due to its momentum resolution, the translation part of a nonsymmorphic symmetry operation plays an important role in determining the spin polarization of photoelectrons, and thus SARPES provides a way to distinguish glide symmetry from mirror symmetry.

Having established the glide-symmetry selection rule, we simulate spin-dependent photoemission from the surface bands of a glide-invariant material KHgSb (010) (Fig.~2) and investigate the manifestation of the glide-symmetry selection rule in this numerical simulation.
The (010) surface of KHgSb hosts four branches of the metallic surface bands with the hourglass-like energy-momentum dispersion inside the bulk band gap~\cite{wang2016hourglass,ezawa2016PRB,Ma2017SciAdv,LiangPRB2017}.
Due to its large SOC, KHgSb is an appropriate test bed for verifying the spin-selection rule imposed by glide symmetry in the spin-dependent photoemission.

  \begin{figure}
  \centering
  \includegraphics[width=1\columnwidth]{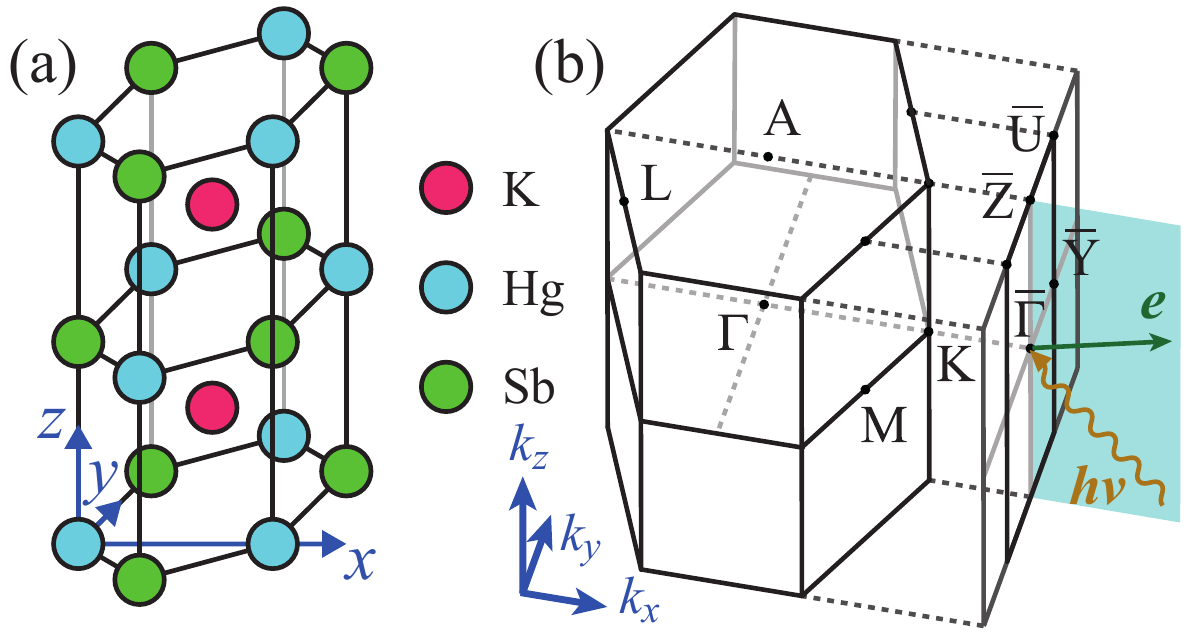}
  \caption{
  (a) The bulk unit cell of KHgSb.
  (b) The bulk and surface Brillouin zones.
  }
  \label{Fig1}
  \end{figure}
  
  \begin{figure}
  \centering
  \includegraphics[width=0.8\columnwidth]{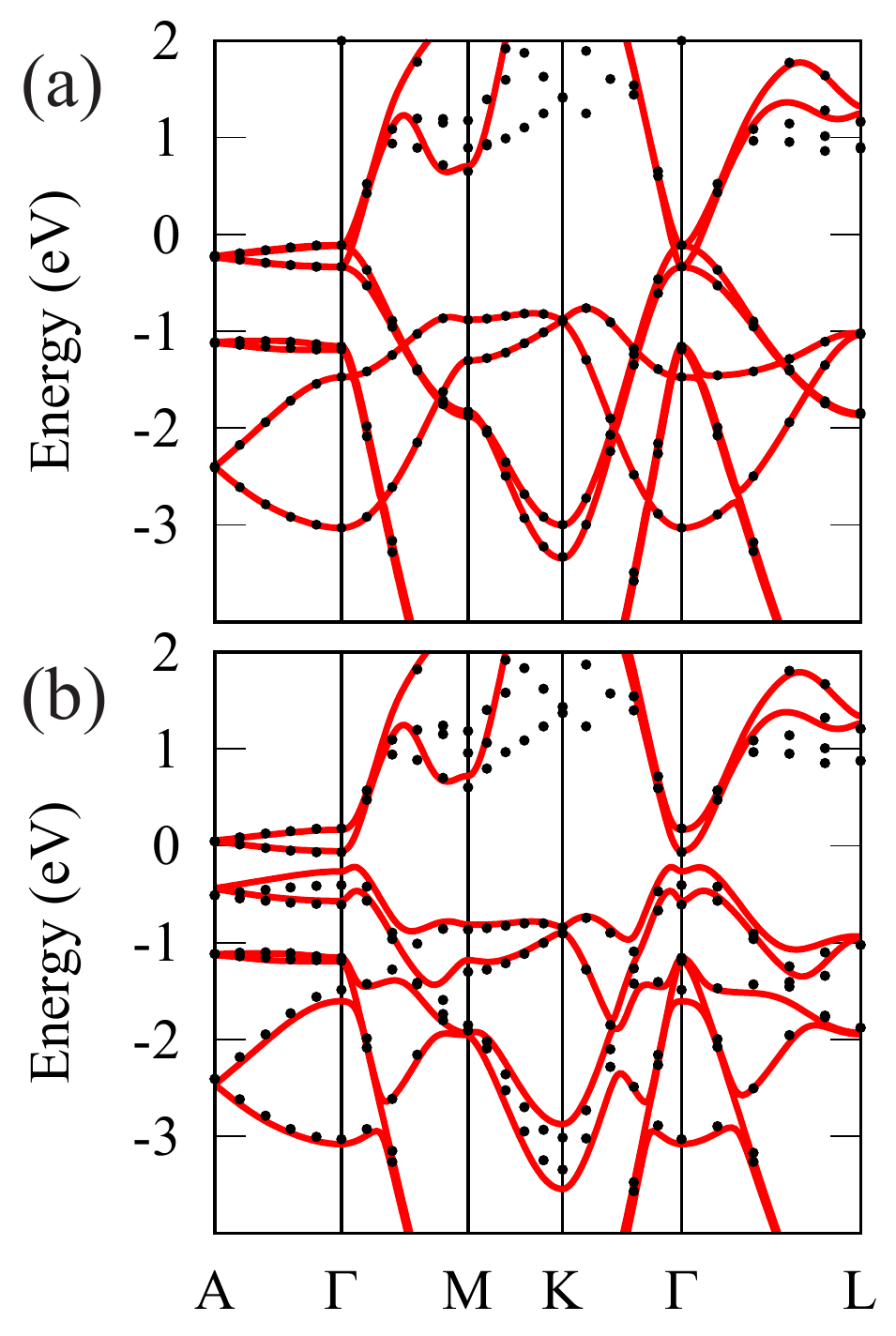}
  \caption{
  The electronic band structure of bulk KHgSb without [(a)] and with [(b)] SOC. Black dots denote the energy eigenvalues obtained from DFT calculations, and solid red curves denote the results obtained from \textit{ab initio} tight binding models containing Hg $6s$ and Sb $5p$ orbitals.
  }
  \label{bulkBand}
  \end{figure}

We constructed the Wannier functions from \textit{ab initio} calculation of the bulk material without considering SOC and extracted the hopping integrals among them (Fig.~3a). Together with an on-site spin-orbit coupling term $\alpha_{\textrm{Sb}}\mathbf{L}\cdot\mathbf{S}/\hbar^2$ with $\alpha_{\textrm{Sb}}=0.56\,\textrm{eV}$ for the $5p$-like orbitals at Sb atoms, the resulting tight-binding Hamiltonian well describes the \textit{ab initio} calculation of the bulk where SOC is fully taken into account near the Fermi energy.  (Fig.~3b).
We then constructed a surface slab containing 60 bulk unit cells along the surface normal direction and calculated the surface band structure (Fig.~4), which shows four hourglass surface bands partially buried in bulk bands.
We remark that our theory applies to any surface bands on a glide-invariant line in the Brillouin zone, not necessarily restricted to hourglass bands.

  \begin{figure}
  \centering
  \includegraphics[width=1\columnwidth]{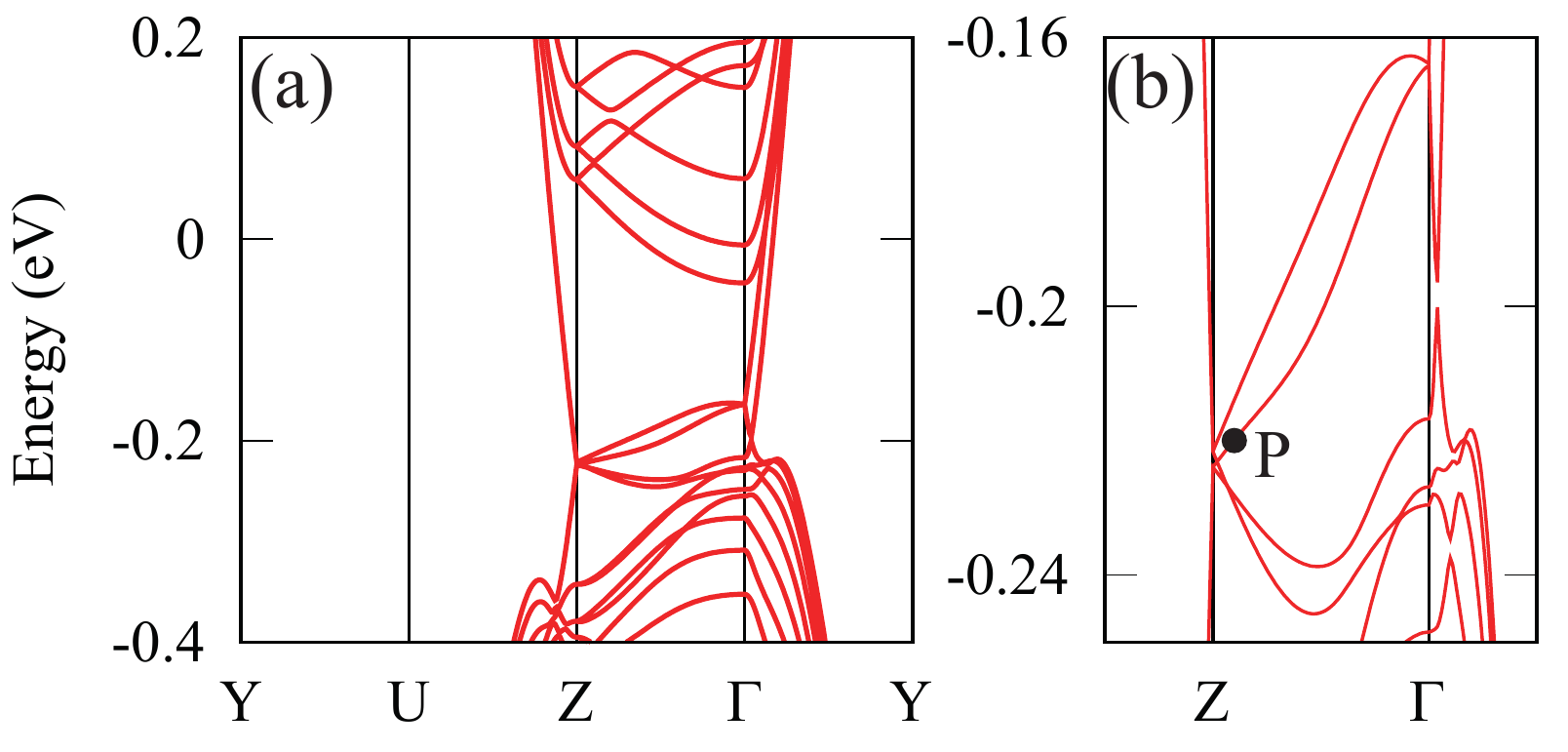}
  \caption{
  (a) The electronic band structure of KHgSb (100).
  (b) Zoom-in of (a) on the hourglass surface bands.
  }
  \label{slabBand}
  \end{figure}

Since the purpose of our paper is not to accurately calculate the photocurrent intensity in ARPES but to demonstrate the selection rule imposed by glide symmetry, it is not necessary to calculate the photoemission final state exactly.
(Nevertheless, as discussed below, a single complex parameter -- the ratio of the dipole matrix elements for $s$- and $p$-polarized light -- determines the relative photoemission intensity and the spin polarization of the photoelectron for an arbitrary light polarization at a $k$-point on the glide-invariant line.)
Therefore, we instead take various combinations of the atomic orbitals as the final state and show that all of these combinations satisfy the spin selection rule shown in Tab.~\ref{tab:summary}.

The final state inside the material can be described as the Bloch sum of atomic orbitals:
\begin{equation}
|f,\sigma\hat{y}\rangle=\sum_{\alpha,\mathbf{R}}{c_{\mathbf{R}\alpha}e^{i\mathbf{k}^f\cdot\mathbf{R}_\alpha-R_{\alpha,z}/2l}|\phi_{\mathbf{R}\alpha}\rangle}.
\label{eq:LCAO}
\end{equation}
Here $\alpha$ denotes the combined index of the atom, orbital and spin in the surface unit cell, $\mathbf{R}_\alpha$ the position of the $\alpha$-th atom in the unit cell displaced by the surface lattice vector $\mathbf{R}$, $|\phi_{\mathbf{R}\alpha}\rangle$ the state representing the $\alpha$-th atomic orbital located at $\mathbf{R}_\alpha$, and $l$ the inelastic mean free path of the final state. We set $l= 10\textrm{~\AA}$ and the final state energy $E^f= 13.6~\mathrm{eV}$ in the calculation below.

  \begin{figure*}
  \centering
  \includegraphics[width=0.85\textwidth]{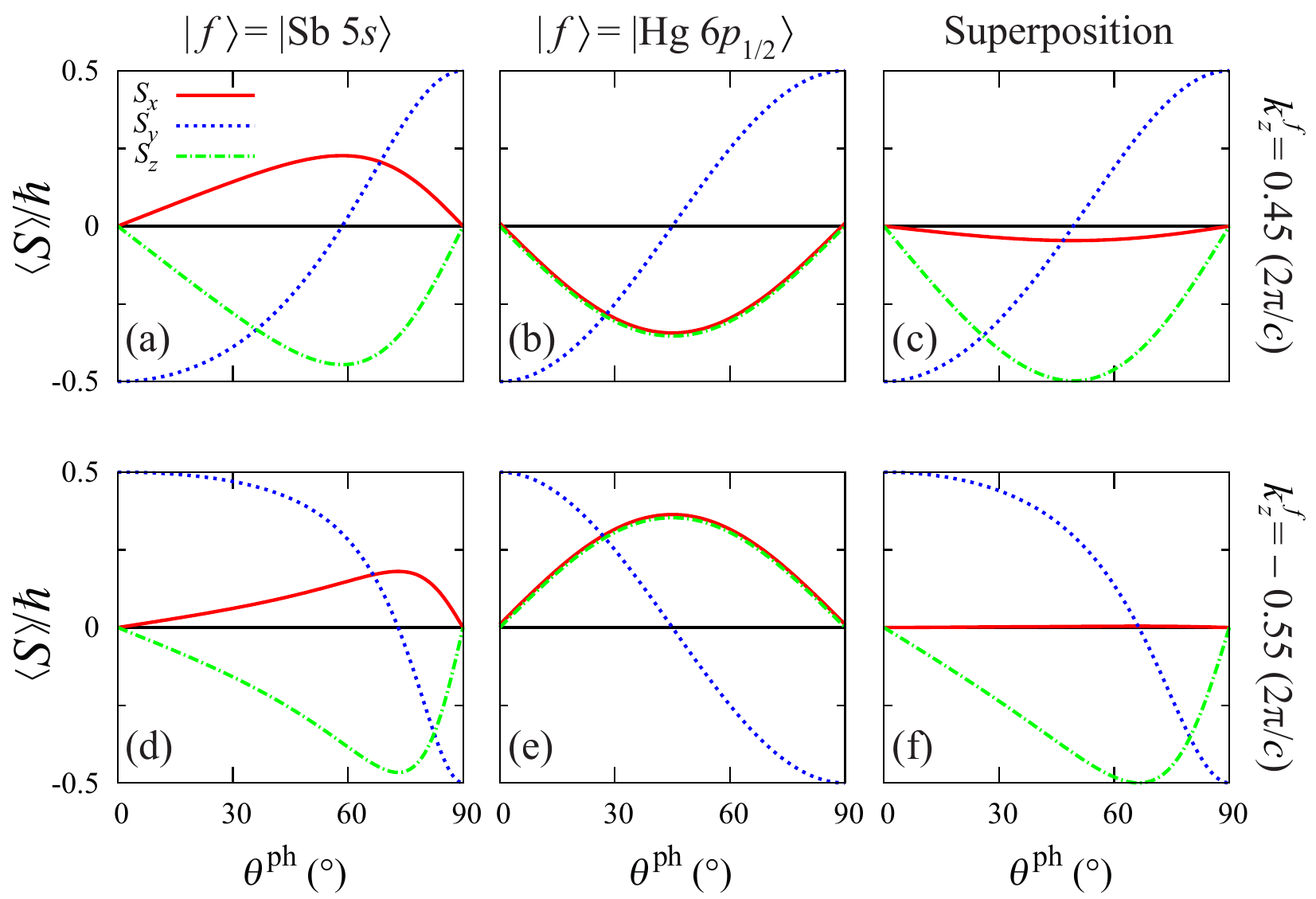}
  \caption{
  Spin polarizations of photoelectrons with $\mathbf{k}^f = (0,0,0.45(2\pi/c))$, ejected from the initial state denoted by P in Fig.~\ref{slabBand} assuming that the final state is composed of (a) Sb 5$s$ orbitals, (b) Hg 6$p$ orbitals with $j=1/2$, and (c) a coherent superposition of both. (d-f) The same quantities as (a-c) for photoelectrons with $\mathbf{k}^f = (0,0,-0.55(2\pi/c))$.
  }
  \label{Fig_spin}
  \end{figure*}

Since the surface state is mainly composed of Sb $5p$ and Hg $6s$ orbitals, the relevant atomic transition is $p\rightarrow s$ transition at Sb and $s \rightarrow p$ transition at Hg by the atomic dipole selection rule.
Having this fact in mind, we consider the following three simplified cases:
(i) the $p\rightarrow s$ transition at Sb is dominant and the effect of SOC on the final state is negligible, in which case $c_{\mathbf{R} \alpha}$ in Eq.~(\ref{eq:LCAO}) is a nonzero constant for $5s$ orbitals at Sb with $s_y=\sigma/2$ and zero otherwise;
(ii) the $s\rightarrow p$ transition at Hg is dominant, the effect of SOC is very strong, and $c_{\mathbf{R} \alpha}$ is a nonzero constant for $5p_{1/2}$ orbitals (i.e., 5$p$ orbitals with total angular momentum $j$=1/2) at Hg with $j_y=\sigma/2$ and zero otherwise;
(iii) $c_{\mathbf{R} \alpha}$ is taken so that the final states in the preceding two cases are superposed.

For those final states, we show in Fig.~5 the spin polarization of the photoelectrons ejected from the surface state denoted by P in Fig.~4, with crystal momentum $\mathbf{k}^i=(0,0,0.45(2\pi/c))$, as a function of the direction of the linear light polarization.
Irrespective of the final state (Fig.~5a-5c), the spin polarization of photoelectrons whose momentum is in the first surface Brillouin zone ($\mathbf{k}^f=0.45(2\pi/c)\hat{z}$) is along $-\hat{y}$ when ejected by $s$-polarized light ($\theta^{\textrm{ph}}=0^\circ$) and is along $+\hat{y}$ when ejected by $p$-polarized light ($\theta^{\textrm{ph}}=90^\circ$).
On the contrary, the spin polarization of photoelectrons whose momentum is in the second Brillouin zone (Fig.~5d-5f, $\mathbf{k}^f=-0.55(2\pi/c)\hat{z}$) is along $+\hat{y}$ when ejected by $s$-polarized light. This result is a direct consequence of the glide symmetry.

When the light is neither $s$- nor $p$-polarized, the photoemission configuration breaks the glide symmetry, i.e., $\mathbf{A}\cdot\mathbf{p}$ is not invariant under the glide reflection. Even in this case, however, any light polarization $\mathbf{A}$ is a linear combination of $s$ and $p$ polarizations $\mathbf{A}_s$ and $\mathbf{A}_p$. For example, when the light polarization is rotated from the $s$ polarization by $\theta^\textrm{ph}$, the photoemission matrix element is $\langle f\sigma|\mathbf{A}\cdot\mathbf{p}|i\rangle=\cos{\theta^{\textrm{ph}}}\langle f\sigma|\mathbf{A}_s\cdot\mathbf{p}|i\rangle+\sin{\theta^{\textrm{ph}}}\langle f\sigma|\mathbf{A}_p\cdot\mathbf{p}|i\rangle$.
(We have suppressed $\hat{y}$ for simplicity.)
Therefore, similar to the mirror-symmetric case~\cite{Kobayashi2017PRB,yaji2017spin}, a single complex parameter, namely, the ratio between the matrix elements for $s$-polarized light $\langle f(-\sigma)|\mathbf{A}_s\cdot\mathbf{p}|i\rangle$ and for $p$-polarized light $\langle f\,\sigma|\mathbf{A}_p\cdot\mathbf{p}|i\rangle$ with $\sigma=(-1)^n \lambda$ determines the relative photoemission intensity and the spin polarization of photoelectrons ejected with arbitrary light polarization.
Conversely, by measuring the spin polarizations of photoelectrons ejected by light with a few different polarizations, we can obtain the ratio of the matrix elements for $s$- and $p$-polarized light, which in turn enables us to predict the SARPES behavior for any other light polarization.
Our finding that the photoemission intensity and the spin polarization are determined from a single complex parameter in glide-symmetric systems when the propagation directions of incident light and photoelectrons are included in the glide plane extends the previous studies of SARPES from mirror-symmetric topological insulators~\cite{Kobayashi2017PRB,yaji2017spin}.

Moreover, we claim that the magnitude of that complex parameter is an indicator of tunability of the spin direction of a possible spin-polarized photocathode using glide- or mirror-symmetric materials~\cite{jozwiak2013photoelectron,park2012spin}. If $\left| \langle f\,\,(-\sigma)|\mathbf{A}_s\cdot\mathbf{p}|i\rangle \right| \ll \left| \langle f\,\,\sigma|\mathbf{A}_p\cdot\mathbf{p}|i\rangle \right|$, then the SARPES behavior is mostly determined by the $p$-polarization component of the light and the spin polarization of photoelectrons will be close to $\sigma \hat{y}$ unless the $p$-polarization component is very small. Even in this case, the photoemission intensity is low due to the small magnitude of $\left| \langle f\,\,(-\sigma)|\mathbf{A}_s\cdot\mathbf{p}|i\rangle \right|$~\cite{Ryoo2016PRB}.
A similar argument holds for the case $\left| \langle f\,\,(-\sigma)|\mathbf{A}_s\cdot\mathbf{p}|i\rangle \right| \gg \left| \langle f\,\,\sigma|\mathbf{A}_p\cdot\mathbf{p}|i\rangle \right|$. Therefore, the magnitudes of the two matrix elements must be similar in order for the spin polarization of photoelectrons to be tuned easily by changing the light polarization.

\section{Conclusions}

In summary, we have studied spin-resolved ARPES of materials with glide symmetry where the propagating directions of the incident light and outgoing electrons are also in the glide plane.
When the light polarization is parallel or perpendicular to the glide plane (say, $zx$ plane), the spin polarization of photoelectrons is $+\hat{y}$ or $-\hat{y}$, which is perpendicular to the glide plane. Whether the spin polarization is $+\hat{y}$ or $-\hat{y}$ is determined by (i) the glide eigenvalue of the initial surface state and (ii) the in-plane momentum of photoelectrons.
Regarding (ii), even if the photoelectrons are ejected from a single initial state by the same light, when the in-plane momentum of photoelectrons change by the smallest surface reciprocal lattice vector, the spin of photoelectrons is reversed.
In particular, when the momentum of photoelectrons are near the first surface zone boundary, the spin polarizations of the photoelectron and the initial state are either parallel or antiparallel to each other depending on whether the momentum of photoelectrons is near one zone boundary or the other.
This momentum-dependent spin selection distinguishes glide symmetry from mirror symmetry.
Not only do these results manifest the nonsymmorphic character of glide reflection, they also show that the spin-resolved ARPES is a powerful tool in studying the topological phases protected by glide symmetry because it directly measures the glide eigenvalue of the initial states.
We also have shown that the spin of photoelectrons are fully controlled by a single complex parameter due to glide symmetry and the magnitude of that complex parameter measures the tuning power of the spin of electrons using light.

\begin{acknowledgments}
This work was supported by the Creative-Pioneering Research Program through Seoul
National University.
\end{acknowledgments}

\bibliography{manuscript}

\begin{thebibliography}{31}%
\makeatletter
\providecommand \@ifxundefined [1]{%
 \@ifx{#1\undefined}
}%
\providecommand \@ifnum [1]{%
 \ifnum #1\expandafter \@firstoftwo
 \else \expandafter \@secondoftwo
 \fi
}%
\providecommand \@ifx [1]{%
 \ifx #1\expandafter \@firstoftwo
 \else \expandafter \@secondoftwo
 \fi
}%
\providecommand \natexlab [1]{#1}%
\providecommand \enquote  [1]{``#1''}%
\providecommand \bibnamefont  [1]{#1}%
\providecommand \bibfnamefont [1]{#1}%
\providecommand \citenamefont [1]{#1}%
\providecommand \href@noop [0]{\@secondoftwo}%
\providecommand \href [0]{\begingroup \@sanitize@url \@href}%
\providecommand \@href[1]{\@@startlink{#1}\@@href}%
\providecommand \@@href[1]{\endgroup#1\@@endlink}%
\providecommand \@sanitize@url [0]{\catcode `\\12\catcode `\$12\catcode
  `\&12\catcode `\#12\catcode `\^12\catcode `\_12\catcode `\%12\relax}%
\providecommand \@@startlink[1]{}%
\providecommand \@@endlink[0]{}%
\providecommand \url  [0]{\begingroup\@sanitize@url \@url }%
\providecommand \@url [1]{\endgroup\@href {#1}{\urlprefix }}%
\providecommand \urlprefix  [0]{URL }%
\providecommand \Eprint [0]{\href }%
\providecommand \doibase [0]{http://dx.doi.org/}%
\providecommand \selectlanguage [0]{\@gobble}%
\providecommand \bibinfo  [0]{\@secondoftwo}%
\providecommand \bibfield  [0]{\@secondoftwo}%
\providecommand \translation [1]{[#1]}%
\providecommand \BibitemOpen [0]{}%
\providecommand \bibitemStop [0]{}%
\providecommand \bibitemNoStop [0]{.\EOS\space}%
\providecommand \EOS [0]{\spacefactor3000\relax}%
\providecommand \BibitemShut  [1]{\csname bibitem#1\endcsname}%
\let\auto@bib@innerbib\@empty
%</preamble>
\bibitem [{\citenamefont {Park}\ and\ \citenamefont
  {Louie}(2012)}]{park2012spin}%
  \BibitemOpen
  \bibfield  {author} {\bibinfo {author} {\bibfnamefont {C.-H.}\ \bibnamefont
  {Park}}\ and\ \bibinfo {author} {\bibfnamefont {S.~G.}\ \bibnamefont
  {Louie}},\ }\bibfield  {title} {\enquote {\bibinfo {title} {Spin polarization
  of photoelectrons from topological insulators},}\ }\href@noop {} {\bibfield
  {journal} {\bibinfo  {journal} {Phys. Rev. Lett.}\ }\textbf {\bibinfo
  {volume} {109}},\ \bibinfo {pages} {097601} (\bibinfo {year}
  {2012})}\BibitemShut {NoStop}%
\bibitem [{\citenamefont {Zhang}\ \emph {et~al.}(2013)\citenamefont {Zhang},
  \citenamefont {Liu},\ and\ \citenamefont {Zhang}}]{zhang2013spin}%
  \BibitemOpen
  \bibfield  {author} {\bibinfo {author} {\bibfnamefont {H.}~\bibnamefont
  {Zhang}}, \bibinfo {author} {\bibfnamefont {C.-X.}\ \bibnamefont {Liu}}, \
  and\ \bibinfo {author} {\bibfnamefont {S.-C.}\ \bibnamefont {Zhang}},\
  }\bibfield  {title} {\enquote {\bibinfo {title} {Spin-orbital texture in
  topological insulators},}\ }\href@noop {} {\bibfield  {journal} {\bibinfo
  {journal} {Phys. Rev. Lett.}\ }\textbf {\bibinfo {volume} {111}},\ \bibinfo
  {pages} {066801} (\bibinfo {year} {2013})}\BibitemShut {NoStop}%
\bibitem [{\citenamefont {Fu}(2009)}]{fu2009hexagonal}%
  \BibitemOpen
  \bibfield  {author} {\bibinfo {author} {\bibfnamefont {L.}~\bibnamefont
  {Fu}},\ }\bibfield  {title} {\enquote {\bibinfo {title} {Hexagonal warping
  effects in the surface states of the topological insulator
  {${\mathrm{Bi}}_{2}{\mathrm{Te}}_{3}$}},}\ }\href {\doibase
  10.1103/PhysRevLett.103.266801} {\bibfield  {journal} {\bibinfo  {journal}
  {Phys. Rev. Lett.}\ }\textbf {\bibinfo {volume} {103}},\ \bibinfo {pages}
  {266801} (\bibinfo {year} {2009})}\BibitemShut {NoStop}%
\bibitem [{\citenamefont {Young}\ and\ \citenamefont
  {Kane}(2015)}]{young2015dirac}%
  \BibitemOpen
  \bibfield  {author} {\bibinfo {author} {\bibfnamefont {S.~M.}\ \bibnamefont
  {Young}}\ and\ \bibinfo {author} {\bibfnamefont {C.~L.}\ \bibnamefont
  {Kane}},\ }\bibfield  {title} {\enquote {\bibinfo {title} {Dirac semimetals
  in two dimensions},}\ }\href@noop {} {\bibfield  {journal} {\bibinfo
  {journal} {Phys. Rev. Lett.}\ }\textbf {\bibinfo {volume} {115}},\ \bibinfo
  {pages} {126803} (\bibinfo {year} {2015})}\BibitemShut {NoStop}%
\bibitem [{\citenamefont {Liu}\ \emph {et~al.}(2014)\citenamefont {Liu},
  \citenamefont {Zhou}, \citenamefont {Zhang}, \citenamefont {Wang},
  \citenamefont {Weng}, \citenamefont {Prabhakaran}, \citenamefont {Mo},
  \citenamefont {Shen}, \citenamefont {Fang}, \citenamefont {Dai},
  \citenamefont {Hussain},\ and\ \citenamefont {Chen}}]{Liu2014discovery}%
  \BibitemOpen
  \bibfield  {author} {\bibinfo {author} {\bibfnamefont {Z.~K.}\ \bibnamefont
  {Liu}}, \bibinfo {author} {\bibfnamefont {B.}~\bibnamefont {Zhou}}, \bibinfo
  {author} {\bibfnamefont {Y.}~\bibnamefont {Zhang}}, \bibinfo {author}
  {\bibfnamefont {Z.~J.}\ \bibnamefont {Wang}}, \bibinfo {author}
  {\bibfnamefont {H.~M.}\ \bibnamefont {Weng}}, \bibinfo {author}
  {\bibfnamefont {D.}~\bibnamefont {Prabhakaran}}, \bibinfo {author}
  {\bibfnamefont {S.-K.}\ \bibnamefont {Mo}}, \bibinfo {author} {\bibfnamefont
  {Z.~X.}\ \bibnamefont {Shen}}, \bibinfo {author} {\bibfnamefont
  {Z.}~\bibnamefont {Fang}}, \bibinfo {author} {\bibfnamefont {X.}~\bibnamefont
  {Dai}}, \bibinfo {author} {\bibfnamefont {Z.}~\bibnamefont {Hussain}}, \ and\
  \bibinfo {author} {\bibfnamefont {Y.~L.}\ \bibnamefont {Chen}},\ }\bibfield
  {title} {\enquote {\bibinfo {title} {Discovery of a three-dimensional
  topological dirac semimetal, {${\mathrm{Na}}_{3}{\mathrm{Bi}}$}},}\ }\href
  {\doibase 10.1126/science.1245085} {\bibfield  {journal} {\bibinfo  {journal}
  {Science}\ }\textbf {\bibinfo {volume} {343}},\ \bibinfo {pages} {864--867}
  (\bibinfo {year} {2014})}\BibitemShut {NoStop}%
\bibitem [{\citenamefont {Jozwiak}\ \emph {et~al.}(2013)\citenamefont
  {Jozwiak}, \citenamefont {Park}, \citenamefont {Gotlieb}, \citenamefont
  {Hwang}, \citenamefont {Lee}, \citenamefont {Louie}, \citenamefont
  {Denlinger}, \citenamefont {Rotundu}, \citenamefont {Birgeneau},
  \citenamefont {Hussain},\ and\ \citenamefont
  {Lanzara}}]{jozwiak2013photoelectron}%
  \BibitemOpen
  \bibfield  {author} {\bibinfo {author} {\bibfnamefont {C.}~\bibnamefont
  {Jozwiak}}, \bibinfo {author} {\bibfnamefont {C.-H.}\ \bibnamefont {Park}},
  \bibinfo {author} {\bibfnamefont {K.}~\bibnamefont {Gotlieb}}, \bibinfo
  {author} {\bibfnamefont {C.}~\bibnamefont {Hwang}}, \bibinfo {author}
  {\bibfnamefont {D.-H.}\ \bibnamefont {Lee}}, \bibinfo {author} {\bibfnamefont
  {S.~G.}\ \bibnamefont {Louie}}, \bibinfo {author} {\bibfnamefont {J.~D.}\
  \bibnamefont {Denlinger}}, \bibinfo {author} {\bibfnamefont {C.~R.}\
  \bibnamefont {Rotundu}}, \bibinfo {author} {\bibfnamefont {R.~J.}\
  \bibnamefont {Birgeneau}}, \bibinfo {author} {\bibfnamefont {Z.}~\bibnamefont
  {Hussain}}, \ and\ \bibinfo {author} {\bibfnamefont {A.}~\bibnamefont
  {Lanzara}},\ }\bibfield  {title} {\enquote {\bibinfo {title} {Photoelectron
  spin-flipping and texture manipulation in a topological insulator},}\
  }\href@noop {} {\bibfield  {journal} {\bibinfo  {journal} {Nat. Phys.}\
  }\textbf {\bibinfo {volume} {9}},\ \bibinfo {pages} {293--298} (\bibinfo
  {year} {2013})}\BibitemShut {NoStop}%
\bibitem [{\citenamefont {Zhu}\ \emph {et~al.}(2013)\citenamefont {Zhu},
  \citenamefont {Veenstra}, \citenamefont {Levy}, \citenamefont {Ubaldini},
  \citenamefont {Syers}, \citenamefont {Butch}, \citenamefont {Paglione},
  \citenamefont {Haverkort}, \citenamefont {Elfimov},\ and\ \citenamefont
  {Damascelli}}]{Damascelli2013PRL}%
  \BibitemOpen
  \bibfield  {author} {\bibinfo {author} {\bibfnamefont {Z.-H.}\ \bibnamefont
  {Zhu}}, \bibinfo {author} {\bibfnamefont {C.~N.}\ \bibnamefont {Veenstra}},
  \bibinfo {author} {\bibfnamefont {G.}~\bibnamefont {Levy}}, \bibinfo {author}
  {\bibfnamefont {A.}~\bibnamefont {Ubaldini}}, \bibinfo {author}
  {\bibfnamefont {P.}~\bibnamefont {Syers}}, \bibinfo {author} {\bibfnamefont
  {N.~P.}\ \bibnamefont {Butch}}, \bibinfo {author} {\bibfnamefont
  {J.}~\bibnamefont {Paglione}}, \bibinfo {author} {\bibfnamefont {M.~W.}\
  \bibnamefont {Haverkort}}, \bibinfo {author} {\bibfnamefont {I.~S.}\
  \bibnamefont {Elfimov}}, \ and\ \bibinfo {author} {\bibfnamefont
  {A.}~\bibnamefont {Damascelli}},\ }\bibfield  {title} {\enquote {\bibinfo
  {title} {Layer-by-layer entangled spin-orbital texture of the topological
  surface state in {${\mathrm{Bi}}_{2}{\mathrm{Se}}_{3}$}},}\ }\href {\doibase
  10.1103/PhysRevLett.110.216401} {\bibfield  {journal} {\bibinfo  {journal}
  {Phys. Rev. Lett.}\ }\textbf {\bibinfo {volume} {110}},\ \bibinfo {pages}
  {216401} (\bibinfo {year} {2013})}\BibitemShut {NoStop}%
\bibitem [{\citenamefont {Zhu}\ \emph {et~al.}(2014)\citenamefont {Zhu},
  \citenamefont {Veenstra}, \citenamefont {Zhdanovich}, \citenamefont
  {Schneider}, \citenamefont {Okuda}, \citenamefont {Miyamoto}, \citenamefont
  {Zhu}, \citenamefont {Namatame}, \citenamefont {Taniguchi}, \citenamefont
  {Haverkort}, \citenamefont {Elfimov},\ and\ \citenamefont
  {Damascelli}}]{Damascelli2014PRL}%
  \BibitemOpen
  \bibfield  {author} {\bibinfo {author} {\bibfnamefont {Z.-H.}\ \bibnamefont
  {Zhu}}, \bibinfo {author} {\bibfnamefont {C.~N.}\ \bibnamefont {Veenstra}},
  \bibinfo {author} {\bibfnamefont {S.}~\bibnamefont {Zhdanovich}}, \bibinfo
  {author} {\bibfnamefont {M.~P.}\ \bibnamefont {Schneider}}, \bibinfo {author}
  {\bibfnamefont {T.}~\bibnamefont {Okuda}}, \bibinfo {author} {\bibfnamefont
  {K.}~\bibnamefont {Miyamoto}}, \bibinfo {author} {\bibfnamefont {S.-Y.}\
  \bibnamefont {Zhu}}, \bibinfo {author} {\bibfnamefont {H.}~\bibnamefont
  {Namatame}}, \bibinfo {author} {\bibfnamefont {M.}~\bibnamefont {Taniguchi}},
  \bibinfo {author} {\bibfnamefont {M.~W.}\ \bibnamefont {Haverkort}}, \bibinfo
  {author} {\bibfnamefont {I.~S.}\ \bibnamefont {Elfimov}}, \ and\ \bibinfo
  {author} {\bibfnamefont {A.}~\bibnamefont {Damascelli}},\ }\bibfield  {title}
  {\enquote {\bibinfo {title} {Photoelectron spin-polarization control in the
  topological insulator {${\mathrm{Bi}}_{2}{\mathrm{Se}}_{3}$}},}\ }\href
  {\doibase 10.1103/PhysRevLett.112.076802} {\bibfield  {journal} {\bibinfo
  {journal} {Phys. Rev. Lett.}\ }\textbf {\bibinfo {volume} {112}},\ \bibinfo
  {pages} {076802} (\bibinfo {year} {2014})}\BibitemShut {NoStop}%
\bibitem [{\citenamefont {Ryoo}\ and\ \citenamefont
  {Park}(2016)}]{Ryoo2016PRB}%
  \BibitemOpen
  \bibfield  {author} {\bibinfo {author} {\bibfnamefont {J.~H.}\ \bibnamefont
  {Ryoo}}\ and\ \bibinfo {author} {\bibfnamefont {C.-H.}\ \bibnamefont
  {Park}},\ }\bibfield  {title} {\enquote {\bibinfo {title} {Spin-conserving
  and reversing photoemission from the surface states of
  {${\mathrm{Bi}}_{2}{\mathrm{Se}}_{3}$ and Au (111)}},}\ }\href {\doibase
  10.1103/PhysRevB.93.085419} {\bibfield  {journal} {\bibinfo  {journal} {Phys.
  Rev. B}\ }\textbf {\bibinfo {volume} {93}},\ \bibinfo {pages} {085419}
  (\bibinfo {year} {2016})}\BibitemShut {NoStop}%
\bibitem [{\citenamefont {Hwang}\ \emph {et~al.}(2011)\citenamefont {Hwang},
  \citenamefont {Park}, \citenamefont {Siegel}, \citenamefont {Fedorov},
  \citenamefont {Louie},\ and\ \citenamefont {Lanzara}}]{Hwang2011PRB}%
  \BibitemOpen
  \bibfield  {author} {\bibinfo {author} {\bibfnamefont {C.}~\bibnamefont
  {Hwang}}, \bibinfo {author} {\bibfnamefont {C.-H.}\ \bibnamefont {Park}},
  \bibinfo {author} {\bibfnamefont {D.~A.}\ \bibnamefont {Siegel}}, \bibinfo
  {author} {\bibfnamefont {A.~V.}\ \bibnamefont {Fedorov}}, \bibinfo {author}
  {\bibfnamefont {S.~G.}\ \bibnamefont {Louie}}, \ and\ \bibinfo {author}
  {\bibfnamefont {A.}~\bibnamefont {Lanzara}},\ }\bibfield  {title} {\enquote
  {\bibinfo {title} {Direct measurement of quantum phases in graphene via
  photoemission spectroscopy},}\ }\href {\doibase 10.1103/PhysRevB.84.125422}
  {\bibfield  {journal} {\bibinfo  {journal} {Phys. Rev. B}\ }\textbf {\bibinfo
  {volume} {84}},\ \bibinfo {pages} {125422} (\bibinfo {year}
  {2011})}\BibitemShut {NoStop}%
\bibitem [{\citenamefont {Kobayashi}\ \emph {et~al.}(2017)\citenamefont
  {Kobayashi}, \citenamefont {Yaji}, \citenamefont {Kuroda},\ and\
  \citenamefont {Komori}}]{Kobayashi2017PRB}%
  \BibitemOpen
  \bibfield  {author} {\bibinfo {author} {\bibfnamefont {K.}~\bibnamefont
  {Kobayashi}}, \bibinfo {author} {\bibfnamefont {K.}~\bibnamefont {Yaji}},
  \bibinfo {author} {\bibfnamefont {K.}~\bibnamefont {Kuroda}}, \ and\ \bibinfo
  {author} {\bibfnamefont {F.}~\bibnamefont {Komori}},\ }\bibfield  {title}
  {\enquote {\bibinfo {title} {Calculation of spin states of photoelectrons
  emitted from spin-polarized surface states of {$\textrm{Bi}(111)$} surfaces
  with a mirror symmetry},}\ }\href {\doibase 10.1103/PhysRevB.95.205436}
  {\bibfield  {journal} {\bibinfo  {journal} {Phys. Rev. B}\ }\textbf {\bibinfo
  {volume} {95}},\ \bibinfo {pages} {205436} (\bibinfo {year}
  {2017})}\BibitemShut {NoStop}%
\bibitem [{\citenamefont {Yaji}\ \emph {et~al.}(2017)\citenamefont {Yaji},
  \citenamefont {Kuroda}, \citenamefont {Toyohisa}, \citenamefont {Harasawa},
  \citenamefont {Ishida}, \citenamefont {Watanabe}, \citenamefont {Chen},
  \citenamefont {Kobayashi}, \citenamefont {Komori},\ and\ \citenamefont
  {Shin}}]{yaji2017spin}%
  \BibitemOpen
  \bibfield  {author} {\bibinfo {author} {\bibfnamefont {K.}~\bibnamefont
  {Yaji}}, \bibinfo {author} {\bibfnamefont {K.}~\bibnamefont {Kuroda}},
  \bibinfo {author} {\bibfnamefont {S.}~\bibnamefont {Toyohisa}}, \bibinfo
  {author} {\bibfnamefont {A.}~\bibnamefont {Harasawa}}, \bibinfo {author}
  {\bibfnamefont {Y.}~\bibnamefont {Ishida}}, \bibinfo {author} {\bibfnamefont
  {S.}~\bibnamefont {Watanabe}}, \bibinfo {author} {\bibfnamefont
  {C.}~\bibnamefont {Chen}}, \bibinfo {author} {\bibfnamefont {K.}~\bibnamefont
  {Kobayashi}}, \bibinfo {author} {\bibfnamefont {F.}~\bibnamefont {Komori}}, \
  and\ \bibinfo {author} {\bibfnamefont {S.}~\bibnamefont {Shin}},\ }\bibfield
  {title} {\enquote {\bibinfo {title} {Spin-dependent quantum interference in
  photoemission process from spin-orbit coupled states},}\ }\href@noop {}
  {\bibfield  {journal} {\bibinfo  {journal} {Nat. Commun.}\ }\textbf {\bibinfo
  {volume} {8}},\ \bibinfo {pages} {14588} (\bibinfo {year}
  {2017})}\BibitemShut {NoStop}%
\bibitem [{\citenamefont {Gotlieb}\ \emph {et~al.}(2017)\citenamefont
  {Gotlieb}, \citenamefont {Li}, \citenamefont {Lin}, \citenamefont {Jozwiak},
  \citenamefont {Ryoo}, \citenamefont {Park}, \citenamefont {Hussain},
  \citenamefont {Louie},\ and\ \citenamefont {Lanzara}}]{Gotlieb2017PRB}%
  \BibitemOpen
  \bibfield  {author} {\bibinfo {author} {\bibfnamefont {K.}~\bibnamefont
  {Gotlieb}}, \bibinfo {author} {\bibfnamefont {Z.}~\bibnamefont {Li}},
  \bibinfo {author} {\bibfnamefont {C.-Y.}\ \bibnamefont {Lin}}, \bibinfo
  {author} {\bibfnamefont {C.}~\bibnamefont {Jozwiak}}, \bibinfo {author}
  {\bibfnamefont {J.~H.}\ \bibnamefont {Ryoo}}, \bibinfo {author}
  {\bibfnamefont {C.-H.}\ \bibnamefont {Park}}, \bibinfo {author}
  {\bibfnamefont {Z.}~\bibnamefont {Hussain}}, \bibinfo {author} {\bibfnamefont
  {S.~G.}\ \bibnamefont {Louie}}, \ and\ \bibinfo {author} {\bibfnamefont
  {A.}~\bibnamefont {Lanzara}},\ }\bibfield  {title} {\enquote {\bibinfo
  {title} {Symmetry rules shaping spin-orbital textures in surface states},}\
  }\href {\doibase 10.1103/PhysRevB.95.245142} {\bibfield  {journal} {\bibinfo
  {journal} {Phys. Rev. B}\ }\textbf {\bibinfo {volume} {95}},\ \bibinfo
  {pages} {245142} (\bibinfo {year} {2017})}\BibitemShut {NoStop}%
\bibitem [{\citenamefont {He}\ \emph {et~al.}(2016)\citenamefont {He},
  \citenamefont {Mion}, \citenamefont {Gao}, \citenamefont {Myers},
  \citenamefont {Arita}, \citenamefont {Shimada}, \citenamefont {Gu},\ and\
  \citenamefont {He}}]{he2016angle}%
  \BibitemOpen
  \bibfield  {author} {\bibinfo {author} {\bibfnamefont {J.}~\bibnamefont
  {He}}, \bibinfo {author} {\bibfnamefont {T.~R.}\ \bibnamefont {Mion}},
  \bibinfo {author} {\bibfnamefont {S.}~\bibnamefont {Gao}}, \bibinfo {author}
  {\bibfnamefont {G.~T.}\ \bibnamefont {Myers}}, \bibinfo {author}
  {\bibfnamefont {M.}~\bibnamefont {Arita}}, \bibinfo {author} {\bibfnamefont
  {K.}~\bibnamefont {Shimada}}, \bibinfo {author} {\bibfnamefont {G.D.}\
  \bibnamefont {Gu}}, \ and\ \bibinfo {author} {\bibfnamefont {R.-H.}\
  \bibnamefont {He}},\ }\bibfield  {title} {\enquote {\bibinfo {title}
  {Angle-resolved photoemission with circularly polarized light in the nodal
  mirror plane of underdoped
  {$\mathrm{Bi}_2\mathrm{Sr}_2\mathrm{Ca}\mathrm{Cu}_2\mathrm{O}_{8+\delta}$}
  superconductor},}\ }\href@noop {} {\bibfield  {journal} {\bibinfo  {journal}
  {Appl. Phys. Lett.}\ }\textbf {\bibinfo {volume} {109}},\ \bibinfo {pages}
  {182601} (\bibinfo {year} {2016})}\BibitemShut {NoStop}%
\bibitem [{\citenamefont {Kuroda}\ \emph {et~al.}(2016)\citenamefont {Kuroda},
  \citenamefont {Yaji}, \citenamefont {Nakayama}, \citenamefont {Harasawa},
  \citenamefont {Ishida}, \citenamefont {Watanabe}, \citenamefont {Chen},
  \citenamefont {Kondo}, \citenamefont {Komori},\ and\ \citenamefont
  {Shin}}]{Kuroda2016PRB}%
  \BibitemOpen
  \bibfield  {author} {\bibinfo {author} {\bibfnamefont {Ke.}\ \bibnamefont
  {Kuroda}}, \bibinfo {author} {\bibfnamefont {K.}~\bibnamefont {Yaji}},
  \bibinfo {author} {\bibfnamefont {M.}~\bibnamefont {Nakayama}}, \bibinfo
  {author} {\bibfnamefont {A.}~\bibnamefont {Harasawa}}, \bibinfo {author}
  {\bibfnamefont {Y.}~\bibnamefont {Ishida}}, \bibinfo {author} {\bibfnamefont
  {S.}~\bibnamefont {Watanabe}}, \bibinfo {author} {\bibfnamefont {C.-T.}\
  \bibnamefont {Chen}}, \bibinfo {author} {\bibfnamefont {T.}~\bibnamefont
  {Kondo}}, \bibinfo {author} {\bibfnamefont {F.}~\bibnamefont {Komori}}, \
  and\ \bibinfo {author} {\bibfnamefont {S.}~\bibnamefont {Shin}},\ }\bibfield
  {title} {\enquote {\bibinfo {title} {Coherent control over three-dimensional
  spin polarization for the spin-orbit coupled surface state of
  {${\mathrm{Bi}}_{2}{\mathrm{Se}}_{3}$}},}\ }\href {\doibase
  10.1103/PhysRevB.94.165162} {\bibfield  {journal} {\bibinfo  {journal} {Phys.
  Rev. B}\ }\textbf {\bibinfo {volume} {94}},\ \bibinfo {pages} {165162}
  (\bibinfo {year} {2016})}\BibitemShut {NoStop}%
\bibitem [{\citenamefont {Pescia}\ \emph {et~al.}(1985)\citenamefont {Pescia},
  \citenamefont {Law}, \citenamefont {Johnson},\ and\ \citenamefont
  {Hughes}}]{pescia1985determination}%
  \BibitemOpen
  \bibfield  {author} {\bibinfo {author} {\bibfnamefont {D.}~\bibnamefont
  {Pescia}}, \bibinfo {author} {\bibfnamefont {A.R.}\ \bibnamefont {Law}},
  \bibinfo {author} {\bibfnamefont {M.T.}\ \bibnamefont {Johnson}}, \ and\
  \bibinfo {author} {\bibfnamefont {H.P.}\ \bibnamefont {Hughes}},\ }\bibfield
  {title} {\enquote {\bibinfo {title} {Determination of observable conduction
  band symmetry in angle-resolved electron spectroscopies: non-symmorphic space
  groups},}\ }\href@noop {} {\bibfield  {journal} {\bibinfo  {journal} {Solid
  State Commun.}\ }\textbf {\bibinfo {volume} {56}},\ \bibinfo {pages}
  {809--812} (\bibinfo {year} {1985})}\BibitemShut {NoStop}%
\bibitem [{\citenamefont {Prince}\ \emph {et~al.}(1986)\citenamefont {Prince},
  \citenamefont {Surman}, \citenamefont {Lindner},\ and\ \citenamefont
  {Bradshaw}}]{prince1986symmetry}%
  \BibitemOpen
  \bibfield  {author} {\bibinfo {author} {\bibfnamefont {K.C.}\ \bibnamefont
  {Prince}}, \bibinfo {author} {\bibfnamefont {M.}~\bibnamefont {Surman}},
  \bibinfo {author} {\bibfnamefont {Th.}\ \bibnamefont {Lindner}}, \ and\
  \bibinfo {author} {\bibfnamefont {A.M.}\ \bibnamefont {Bradshaw}},\
  }\bibfield  {title} {\enquote {\bibinfo {title} {The symmetry-based
  constraints in angle-resolved photoemission from structures belonging to
  non-symmorphic space groups: {p$(2\times2)-{\mathrm{C}}/{\mathrm{Ni}}$
  (100)}},}\ }\href@noop {} {\bibfield  {journal} {\bibinfo  {journal} {Solid
  State Commun.}\ }\textbf {\bibinfo {volume} {59}},\ \bibinfo {pages} {71--75}
  (\bibinfo {year} {1986})}\BibitemShut {NoStop}%
\bibitem [{\citenamefont {Arpiainen}\ and\ \citenamefont
  {Lindroos}(2006)}]{Arpiainen2006PRL_BSCCO}%
  \BibitemOpen
  \bibfield  {author} {\bibinfo {author} {\bibfnamefont {V.}~\bibnamefont
  {Arpiainen}}\ and\ \bibinfo {author} {\bibfnamefont {M.}~\bibnamefont
  {Lindroos}},\ }\bibfield  {title} {\enquote {\bibinfo {title} {Effect of
  symmetry distortions on photoelectron selection rules and spectra of
  {${\mathrm{Bi}}_{2}{\mathrm{Sr}}_{2}{\mathrm{CaCu}}_{2}{\mathrm{O}}_{8+\ensuremath{\delta}}$}},}\
  }\href {\doibase 10.1103/PhysRevLett.97.037601} {\bibfield  {journal}
  {\bibinfo  {journal} {Phys. Rev. Lett.}\ }\textbf {\bibinfo {volume} {97}},\
  \bibinfo {pages} {037601} (\bibinfo {year} {2006})}\BibitemShut {NoStop}%
\bibitem [{\citenamefont {Fang}\ \emph {et~al.}(2015)\citenamefont {Fang},
  \citenamefont {Chen}, \citenamefont {Kee},\ and\ \citenamefont
  {Fu}}]{Fu2015PRL}%
  \BibitemOpen
  \bibfield  {author} {\bibinfo {author} {\bibfnamefont {C.}~\bibnamefont
  {Fang}}, \bibinfo {author} {\bibfnamefont {Y.}~\bibnamefont {Chen}}, \bibinfo
  {author} {\bibfnamefont {H.-Y.}\ \bibnamefont {Kee}}, \ and\ \bibinfo
  {author} {\bibfnamefont {L.}~\bibnamefont {Fu}},\ }\bibfield  {title}
  {\enquote {\bibinfo {title} {Topological nodal line semimetals with and
  without spin-orbital coupling},}\ }\href {\doibase
  10.1103/PhysRevB.92.081201} {\bibfield  {journal} {\bibinfo  {journal} {Phys.
  Rev. B}\ }\textbf {\bibinfo {volume} {92}},\ \bibinfo {pages} {081201}
  (\bibinfo {year} {2015})}\BibitemShut {NoStop}%
\bibitem [{\citenamefont {Bzdu{\v{s}}ek}\ \emph {et~al.}(2016)\citenamefont
  {Bzdu{\v{s}}ek}, \citenamefont {Wu}, \citenamefont {R{\"u}egg}, \citenamefont
  {Sigrist},\ and\ \citenamefont {Soluyanov}}]{bzduvsek2016nodal}%
  \BibitemOpen
  \bibfield  {author} {\bibinfo {author} {\bibfnamefont {T.}~\bibnamefont
  {Bzdu{\v{s}}ek}}, \bibinfo {author} {\bibfnamefont {Q.}~\bibnamefont {Wu}},
  \bibinfo {author} {\bibfnamefont {A.}~\bibnamefont {R{\"u}egg}}, \bibinfo
  {author} {\bibfnamefont {M.}~\bibnamefont {Sigrist}}, \ and\ \bibinfo
  {author} {\bibfnamefont {A.~A.}\ \bibnamefont {Soluyanov}},\ }\bibfield
  {title} {\enquote {\bibinfo {title} {Nodal-chain metals},}\ }\href@noop {}
  {\bibfield  {journal} {\bibinfo  {journal} {Nature}\ }\textbf {\bibinfo
  {volume} {538}},\ \bibinfo {pages} {75} (\bibinfo {year} {2016})}\BibitemShut
  {NoStop}%
\bibitem [{\citenamefont {Shao}\ \emph {et~al.}(2018)\citenamefont {Shao},
  \citenamefont {Chen}, \citenamefont {Gu}, \citenamefont {Guo}, \citenamefont
  {Lu}, \citenamefont {Sun}, \citenamefont {Sheng},\ and\ \citenamefont
  {Xing}}]{shao2018nonsymmorphic}%
  \BibitemOpen
  \bibfield  {author} {\bibinfo {author} {\bibfnamefont {D.}~\bibnamefont
  {Shao}}, \bibinfo {author} {\bibfnamefont {T.}~\bibnamefont {Chen}}, \bibinfo
  {author} {\bibfnamefont {Q.}~\bibnamefont {Gu}}, \bibinfo {author}
  {\bibfnamefont {Z.}~\bibnamefont {Guo}}, \bibinfo {author} {\bibfnamefont
  {P.}~\bibnamefont {Lu}}, \bibinfo {author} {\bibfnamefont {J.}~\bibnamefont
  {Sun}}, \bibinfo {author} {\bibfnamefont {L.}~\bibnamefont {Sheng}}, \ and\
  \bibinfo {author} {\bibfnamefont {D.}~\bibnamefont {Xing}},\ }\bibfield
  {title} {\enquote {\bibinfo {title} {Nonsymmorphic symmetry protected
  node-line semimetal in the trigonal {${\mathrm{Y}}{\mathrm{H}}_{3}$}},}\
  }\href@noop {} {\bibfield  {journal} {\bibinfo  {journal} {Sci. Rep.}\
  }\textbf {\bibinfo {volume} {8}},\ \bibinfo {pages} {1467} (\bibinfo {year}
  {2018})}\BibitemShut {NoStop}%
\bibitem [{\citenamefont {Ehlen}\ \emph {et~al.}(2018)\citenamefont {Ehlen},
  \citenamefont {Sanna}, \citenamefont {Senkovskiy}, \citenamefont {Petaccia},
  \citenamefont {Fedorov}, \citenamefont {Profeta},\ and\ \citenamefont
  {Gr\"uneis}}]{Ehlen2018PRB}%
  \BibitemOpen
  \bibfield  {author} {\bibinfo {author} {\bibfnamefont {N.}~\bibnamefont
  {Ehlen}}, \bibinfo {author} {\bibfnamefont {A.}~\bibnamefont {Sanna}},
  \bibinfo {author} {\bibfnamefont {B.~V.}\ \bibnamefont {Senkovskiy}},
  \bibinfo {author} {\bibfnamefont {L.}~\bibnamefont {Petaccia}}, \bibinfo
  {author} {\bibfnamefont {A.~V.}\ \bibnamefont {Fedorov}}, \bibinfo {author}
  {\bibfnamefont {G.}~\bibnamefont {Profeta}}, \ and\ \bibinfo {author}
  {\bibfnamefont {A.}~\bibnamefont {Gr\"uneis}},\ }\bibfield  {title} {\enquote
  {\bibinfo {title} {Direct observation of a surface resonance state and
  surface band inversion control in black phosphorus},}\ }\href {\doibase
  10.1103/PhysRevB.97.045143} {\bibfield  {journal} {\bibinfo  {journal} {Phys.
  Rev. B}\ }\textbf {\bibinfo {volume} {97}},\ \bibinfo {pages} {045143}
  (\bibinfo {year} {2018})}\BibitemShut {NoStop}%
\bibitem [{\citenamefont {Wang}\ \emph {et~al.}(2016)\citenamefont {Wang},
  \citenamefont {Alexandradinata}, \citenamefont {Cava},\ and\ \citenamefont
  {Bernevig}}]{wang2016hourglass}%
  \BibitemOpen
  \bibfield  {author} {\bibinfo {author} {\bibfnamefont {Z.}~\bibnamefont
  {Wang}}, \bibinfo {author} {\bibfnamefont {A.}~\bibnamefont
  {Alexandradinata}}, \bibinfo {author} {\bibfnamefont {R.~J}\ \bibnamefont
  {Cava}}, \ and\ \bibinfo {author} {\bibfnamefont {B.~A.}\ \bibnamefont
  {Bernevig}},\ }\bibfield  {title} {\enquote {\bibinfo {title} {Hourglass
  fermions},}\ }\href@noop {} {\bibfield  {journal} {\bibinfo  {journal}
  {Nature}\ }\textbf {\bibinfo {volume} {532}},\ \bibinfo {pages} {189}
  (\bibinfo {year} {2016})}\BibitemShut {NoStop}%
\bibitem [{\citenamefont {Ezawa}(2016)}]{ezawa2016PRB}%
  \BibitemOpen
  \bibfield  {author} {\bibinfo {author} {\bibfnamefont {M.}~\bibnamefont
  {Ezawa}},\ }\bibfield  {title} {\enquote {\bibinfo {title} {Hourglass fermion
  surface states in stacked topological insulators with nonsymmorphic
  symmetry},}\ }\href {\doibase 10.1103/PhysRevB.94.155148} {\bibfield
  {journal} {\bibinfo  {journal} {Phys. Rev. B}\ }\textbf {\bibinfo {volume}
  {94}},\ \bibinfo {pages} {155148} (\bibinfo {year} {2016})}\BibitemShut
  {NoStop}%
\bibitem [{\citenamefont {Giannozzi}\ \emph {et~al.}(2009)\citenamefont
  {Giannozzi}, \citenamefont {Baroni}, \citenamefont {Bonini}, \citenamefont
  {Calandra}, \citenamefont {Car}, \citenamefont {Cavazzoni}, \citenamefont
  {Ceresoli}, \citenamefont {Chiarotti}, \citenamefont {Cococcioni},
  \citenamefont {Dabo}, \citenamefont {Dal~Corso}, \citenamefont
  {de~Gironcoli}, \citenamefont {Fabris}, \citenamefont {Fratesi},
  \citenamefont {Gebauer}, \citenamefont {Gerstmann}, \citenamefont
  {Gougoussis}, \citenamefont {Kokalj}, \citenamefont {Lazzeri}, \citenamefont
  {Martin-Samos}, \citenamefont {Marzari}, \citenamefont {Mauri}, \citenamefont
  {Mazzarello}, \citenamefont {Paolini}, \citenamefont {Pasquarello},
  \citenamefont {Paulatto}, \citenamefont {Sbraccia}, \citenamefont {Scandolo},
  \citenamefont {Sclauzero}, \citenamefont {Seitsonen}, \citenamefont
  {Smogunov}, \citenamefont {Umari},\ and\ \citenamefont
  {Wentzcovitch}}]{giannozzi2009quantum}%
  \BibitemOpen
  \bibfield  {author} {\bibinfo {author} {\bibfnamefont {P.}~\bibnamefont
  {Giannozzi}}, \bibinfo {author} {\bibfnamefont {S.}~\bibnamefont {Baroni}},
  \bibinfo {author} {\bibfnamefont {N.}~\bibnamefont {Bonini}}, \bibinfo
  {author} {\bibfnamefont {M.}~\bibnamefont {Calandra}}, \bibinfo {author}
  {\bibfnamefont {R.}~\bibnamefont {Car}}, \bibinfo {author} {\bibfnamefont
  {C.}~\bibnamefont {Cavazzoni}}, \bibinfo {author} {\bibfnamefont
  {D.}~\bibnamefont {Ceresoli}}, \bibinfo {author} {\bibfnamefont {G.~L.}\
  \bibnamefont {Chiarotti}}, \bibinfo {author} {\bibfnamefont {M.}~\bibnamefont
  {Cococcioni}}, \bibinfo {author} {\bibfnamefont {I.}~\bibnamefont {Dabo}},
  \bibinfo {author} {\bibfnamefont {A.}~\bibnamefont {Dal~Corso}}, \bibinfo
  {author} {\bibfnamefont {S.}~\bibnamefont {de~Gironcoli}}, \bibinfo {author}
  {\bibfnamefont {S.}~\bibnamefont {Fabris}}, \bibinfo {author} {\bibfnamefont
  {G.}~\bibnamefont {Fratesi}}, \bibinfo {author} {\bibfnamefont
  {R.}~\bibnamefont {Gebauer}}, \bibinfo {author} {\bibfnamefont
  {U.}~\bibnamefont {Gerstmann}}, \bibinfo {author} {\bibfnamefont
  {C.}~\bibnamefont {Gougoussis}}, \bibinfo {author} {\bibfnamefont
  {A.}~\bibnamefont {Kokalj}}, \bibinfo {author} {\bibfnamefont
  {M.}~\bibnamefont {Lazzeri}}, \bibinfo {author} {\bibfnamefont
  {L.}~\bibnamefont {Martin-Samos}}, \bibinfo {author} {\bibfnamefont
  {N.}~\bibnamefont {Marzari}}, \bibinfo {author} {\bibfnamefont
  {F.}~\bibnamefont {Mauri}}, \bibinfo {author} {\bibfnamefont
  {R.}~\bibnamefont {Mazzarello}}, \bibinfo {author} {\bibfnamefont
  {S.}~\bibnamefont {Paolini}}, \bibinfo {author} {\bibfnamefont
  {A.}~\bibnamefont {Pasquarello}}, \bibinfo {author} {\bibfnamefont
  {L.}~\bibnamefont {Paulatto}}, \bibinfo {author} {\bibfnamefont
  {C.}~\bibnamefont {Sbraccia}}, \bibinfo {author} {\bibfnamefont
  {S.}~\bibnamefont {Scandolo}}, \bibinfo {author} {\bibfnamefont
  {G.}~\bibnamefont {Sclauzero}}, \bibinfo {author} {\bibfnamefont {A.~P.}\
  \bibnamefont {Seitsonen}}, \bibinfo {author} {\bibfnamefont {A.}~\bibnamefont
  {Smogunov}}, \bibinfo {author} {\bibfnamefont {P.}~\bibnamefont {Umari}}, \
  and\ \bibinfo {author} {\bibfnamefont {R.M.}\ \bibnamefont {Wentzcovitch}},\
  }\bibfield  {title} {\enquote {\bibinfo {title} {{QUANTUM ESPRESSO}: a
  modular and open-source software project for quantum simulations of
  materials},}\ }\href@noop {} {\bibfield  {journal} {\bibinfo  {journal} {J.
  Phys.: Condens. Matter}\ }\textbf {\bibinfo {volume} {21}},\ \bibinfo {pages}
  {395502} (\bibinfo {year} {2009})}\BibitemShut {NoStop}%
\bibitem [{\citenamefont {Perdew}\ \emph {et~al.}(2008)\citenamefont {Perdew},
  \citenamefont {Ruzsinszky}, \citenamefont {Csonka}, \citenamefont {Vydrov},
  \citenamefont {Scuseria}, \citenamefont {Constantin}, \citenamefont {Zhou},\
  and\ \citenamefont {Burke}}]{PerdewPRL2008}%
  \BibitemOpen
  \bibfield  {author} {\bibinfo {author} {\bibfnamefont {J.~P.}\ \bibnamefont
  {Perdew}}, \bibinfo {author} {\bibfnamefont {A.}~\bibnamefont {Ruzsinszky}},
  \bibinfo {author} {\bibfnamefont {G.~I.}\ \bibnamefont {Csonka}}, \bibinfo
  {author} {\bibfnamefont {O.~A.}\ \bibnamefont {Vydrov}}, \bibinfo {author}
  {\bibfnamefont {G.~E.}\ \bibnamefont {Scuseria}}, \bibinfo {author}
  {\bibfnamefont {L.~A.}\ \bibnamefont {Constantin}}, \bibinfo {author}
  {\bibfnamefont {X.}~\bibnamefont {Zhou}}, \ and\ \bibinfo {author}
  {\bibfnamefont {K.}~\bibnamefont {Burke}},\ }\bibfield  {title} {\enquote
  {\bibinfo {title} {Restoring the density-gradient expansion for exchange in
  solids and surfaces},}\ }\href {\doibase 10.1103/PhysRevLett.100.136406}
  {\bibfield  {journal} {\bibinfo  {journal} {Phys. Rev. Lett.}\ }\textbf
  {\bibinfo {volume} {100}},\ \bibinfo {pages} {136406} (\bibinfo {year}
  {2008})}\BibitemShut {NoStop}%
\bibitem [{\citenamefont {Mostofi}\ \emph {et~al.}(2014)\citenamefont
  {Mostofi}, \citenamefont {Yates}, \citenamefont {Pizzi}, \citenamefont {Lee},
  \citenamefont {Souza}, \citenamefont {Vanderbilt},\ and\ \citenamefont
  {Marzari}}]{mostofi2014updated}%
  \BibitemOpen
  \bibfield  {author} {\bibinfo {author} {\bibfnamefont {A.~A.}\ \bibnamefont
  {Mostofi}}, \bibinfo {author} {\bibfnamefont {J.~R.}\ \bibnamefont {Yates}},
  \bibinfo {author} {\bibfnamefont {G.}~\bibnamefont {Pizzi}}, \bibinfo
  {author} {\bibfnamefont {Y.-S.}\ \bibnamefont {Lee}}, \bibinfo {author}
  {\bibfnamefont {I.}~\bibnamefont {Souza}}, \bibinfo {author} {\bibfnamefont
  {Da.}\ \bibnamefont {Vanderbilt}}, \ and\ \bibinfo {author} {\bibfnamefont
  {N.}~\bibnamefont {Marzari}},\ }\bibfield  {title} {\enquote {\bibinfo
  {title} {An updated version of wannier90: A tool for obtaining
  maximally-localised wannier functions},}\ }\href@noop {} {\bibfield
  {journal} {\bibinfo  {journal} {Comput. Phys. Commun.}\ }\textbf {\bibinfo
  {volume} {185}},\ \bibinfo {pages} {2309--2310} (\bibinfo {year}
  {2014})}\BibitemShut {NoStop}%
\bibitem [{\citenamefont {Kleiner}(1966)}]{Kleiner1966PR}%
  \BibitemOpen
  \bibfield  {author} {\bibinfo {author} {\bibfnamefont {W.~H.}\ \bibnamefont
  {Kleiner}},\ }\bibfield  {title} {\enquote {\bibinfo {title} {Space-time
  symmetry of transport coefficients},}\ }\href {\doibase
  10.1103/PhysRev.142.318} {\bibfield  {journal} {\bibinfo  {journal} {Phys.
  Rev.}\ }\textbf {\bibinfo {volume} {142}},\ \bibinfo {pages} {318--326}
  (\bibinfo {year} {1966})}\BibitemShut {NoStop}%
\bibitem [{com()}]{comment_kubo}%
  \BibitemOpen
  \href@noop {} {}\bibinfo {note} {For example, consider the Kubo formula for
  the optical conductivity tensor: $\sigma_{\alpha\beta}(\omega)\propto
  \int_0^\beta d\tau \int_0^\infty dt {\left\langle
  [J_\alpha(-i\hbar\tau),J_\beta(t)] \right\rangle}$, where the angular bracket
  $\langle \cdots \rangle$ denotes the thermal ensemble average at a given
  temperature and $\mathbf{J}$ the total current operator proportional to the
  sum of the momenta of all electrons inside the material. Since $\mathbf{J}$
  is a vector operator invariant with respect to translation, if the crystal
  has a symmetry $\mathbf{r}\mapsto A\mathbf{r}+\mathbf{b}$ for some orthogonal
  matrix $A$, then $A^{-1}\sigma(\omega)A=\sigma(\omega)$. This proves that the
  optical conductivity is invariant under the action of the point group, which
  is the quotient of the space group by its translational subgroup. In
  particular, mirror and glide symmetries impose the same constraint on
  $\sigma(\omega)$ as long as their invariant plane is the same.}\BibitemShut
  {Stop}%
\bibitem [{\citenamefont {Ma}\ \emph {et~al.}(2017)\citenamefont {Ma},
  \citenamefont {Lv}, \citenamefont {Wang}, \citenamefont {Nie}, \citenamefont
  {Wang}, \citenamefont {Kong}, \citenamefont {Huang}, \citenamefont {Richard},
  \citenamefont {Zhang}, \citenamefont {Yaji}, \citenamefont {Kuroda},
  \citenamefont {Shin}, \citenamefont {Weng}, \citenamefont {Bernevig},
  \citenamefont {Shi}, \citenamefont {Qian},\ and\ \citenamefont
  {Ding}}]{Ma2017SciAdv}%
  \BibitemOpen
  \bibfield  {author} {\bibinfo {author} {\bibfnamefont {C.}~\bibnamefont {Ma},
  \bibfnamefont {J.and~Yi}}, \bibinfo {author} {\bibfnamefont {B.}~\bibnamefont
  {Lv}}, \bibinfo {author} {\bibfnamefont {Z.}~\bibnamefont {Wang}}, \bibinfo
  {author} {\bibfnamefont {S.}~\bibnamefont {Nie}}, \bibinfo {author}
  {\bibfnamefont {L.}~\bibnamefont {Wang}}, \bibinfo {author} {\bibfnamefont
  {L.}~\bibnamefont {Kong}}, \bibinfo {author} {\bibfnamefont {Y.}~\bibnamefont
  {Huang}}, \bibinfo {author} {\bibfnamefont {P.}~\bibnamefont {Richard}},
  \bibinfo {author} {\bibfnamefont {P.}~\bibnamefont {Zhang}}, \bibinfo
  {author} {\bibfnamefont {K.}~\bibnamefont {Yaji}}, \bibinfo {author}
  {\bibfnamefont {K.}~\bibnamefont {Kuroda}}, \bibinfo {author} {\bibfnamefont
  {S.}~\bibnamefont {Shin}}, \bibinfo {author} {\bibfnamefont {H.}~\bibnamefont
  {Weng}}, \bibinfo {author} {\bibfnamefont {B.~A.}\ \bibnamefont {Bernevig}},
  \bibinfo {author} {\bibfnamefont {Y.}~\bibnamefont {Shi}}, \bibinfo {author}
  {\bibfnamefont {T.}~\bibnamefont {Qian}}, \ and\ \bibinfo {author}
  {\bibfnamefont {H.}~\bibnamefont {Ding}},\ }\bibfield  {title} {\enquote
  {\bibinfo {title} {Experimental evidence of hourglass fermion in the
  candidate nonsymmorphic topological insulator {KHgSb}},}\ }\href
  {http://advances.sciencemag.org/content/3/5/e1602415} {\bibfield  {journal}
  {\bibinfo  {journal} {Sci. Adv.}\ }\textbf {\bibinfo {volume} {3}} (\bibinfo
  {year} {2017})}\BibitemShut {NoStop}%
\bibitem [{\citenamefont {Liang}\ \emph {et~al.}(2017)\citenamefont {Liang},
  \citenamefont {Jiang}, \citenamefont {Wang}, \citenamefont {Sun},
  \citenamefont {Kumar}, \citenamefont {Shekhar}, \citenamefont {Chen},
  \citenamefont {Peng}, \citenamefont {Wang}, \citenamefont {Xu}, \citenamefont
  {Yang}, \citenamefont {Cui}, \citenamefont {Hong}, \citenamefont {Xia},
  \citenamefont {Mo}, \citenamefont {Gao}, \citenamefont {Zhou}, \citenamefont
  {Yang}, \citenamefont {Felser}, \citenamefont {Yan}, \citenamefont {Liu},\
  and\ \citenamefont {Chen}}]{LiangPRB2017}%
  \BibitemOpen
  \bibfield  {author} {\bibinfo {author} {\bibfnamefont {A.~J.}\ \bibnamefont
  {Liang}}, \bibinfo {author} {\bibfnamefont {J.}~\bibnamefont {Jiang}},
  \bibinfo {author} {\bibfnamefont {M.~X.}\ \bibnamefont {Wang}}, \bibinfo
  {author} {\bibfnamefont {Y.}~\bibnamefont {Sun}}, \bibinfo {author}
  {\bibfnamefont {N.}~\bibnamefont {Kumar}}, \bibinfo {author} {\bibfnamefont
  {C.}~\bibnamefont {Shekhar}}, \bibinfo {author} {\bibfnamefont
  {C.}~\bibnamefont {Chen}}, \bibinfo {author} {\bibfnamefont {H.}~\bibnamefont
  {Peng}}, \bibinfo {author} {\bibfnamefont {C.~W.}\ \bibnamefont {Wang}},
  \bibinfo {author} {\bibfnamefont {X.}~\bibnamefont {Xu}}, \bibinfo {author}
  {\bibfnamefont {H.~F.}\ \bibnamefont {Yang}}, \bibinfo {author}
  {\bibfnamefont {S.~T.}\ \bibnamefont {Cui}}, \bibinfo {author} {\bibfnamefont
  {G.~H.}\ \bibnamefont {Hong}}, \bibinfo {author} {\bibfnamefont {Y.-Y.}\
  \bibnamefont {Xia}}, \bibinfo {author} {\bibfnamefont {S.-K.}\ \bibnamefont
  {Mo}}, \bibinfo {author} {\bibfnamefont {Q.}~\bibnamefont {Gao}}, \bibinfo
  {author} {\bibfnamefont {X.~J.}\ \bibnamefont {Zhou}}, \bibinfo {author}
  {\bibfnamefont {L.~X.}\ \bibnamefont {Yang}}, \bibinfo {author}
  {\bibfnamefont {C.}~\bibnamefont {Felser}}, \bibinfo {author} {\bibfnamefont
  {B.~H.}\ \bibnamefont {Yan}}, \bibinfo {author} {\bibfnamefont {Z.~K.}\
  \bibnamefont {Liu}}, \ and\ \bibinfo {author} {\bibfnamefont {Y.~L.}\
  \bibnamefont {Chen}},\ }\bibfield  {title} {\enquote {\bibinfo {title}
  {Observation of the topological surface state in the nonsymmorphic
  topological insulator {KHgSb}},}\ }\href {\doibase
  10.1103/PhysRevB.96.165143} {\bibfield  {journal} {\bibinfo  {journal} {Phys.
  Rev. B}\ }\textbf {\bibinfo {volume} {96}},\ \bibinfo {pages} {165143}
  (\bibinfo {year} {2017})}\BibitemShut {NoStop}%
\end{thebibliography}%

\end{document}